\documentclass[11pt,fullpage]{article}
   \PassOptionsToPackage{numbers, compress}{natbib}

  \usepackage[final]{neurips_2019}

\usepackage[utf8]{inputenc} 
\usepackage[T1]{fontenc}    
\usepackage{hyperref}       
\usepackage{url}            
\usepackage{booktabs}       
\usepackage{amsfonts}       
\usepackage{nicefrac}       
\usepackage{microtype}      
\usepackage{amsmath,graphicx}

\DeclareMathOperator*{\argmin}{arg\,min}

\title{Improving Limited Angle CT Reconstruction\\ with a Robust GAN Prior}

\author{
 Rushil Anirudh\\
  Lawrence Livermore National Laboratory \\
   \texttt{anirudh1@llnl.gov} \\
   \And
  Hyojin Kim \\
  Lawrence Livermore National Laboratory \\
   \texttt{kim63@llnl.gov} \\
   \And
  Jayaraman J. Thiagarajan \\
  Lawrence Livermore National Laboratory \\
   \texttt{jayaramanthi1@llnl.gov} \\
   \And
   K. Aditya Mohan \\
  Lawrence Livermore National Laboratory \\
   \texttt{mohan3@llnl.gov} \\
   \And
    Kyle M. Champley \\
  Lawrence Livermore National Laboratory \\
   \texttt{champley1@llnl.gov} \\
}

\begin{document}

\maketitle

\begin{abstract}
  Limited angle CT reconstruction is an under-determined linear inverse problem that requires appropriate regularization techniques to be solved. In this work we study how pre-trained generative adversarial networks (GANs) can be used to \emph{clean} noisy, highly artifact laden reconstructions from conventional techniques, by effectively projecting onto the inferred image manifold. In particular, we use a robust version of the popularly used GAN prior for inverse problems, based on a recent technique called corruption mimicking, that significantly improves the reconstruction quality. The proposed approach operates in the image space directly, as a result of which it does not need to be trained or require access to the measurement model, is scanner agnostic, and can work over a wide range of sensing scenarios.
 \let\thefootnote\relax \footnotetext{This work was performed under the auspices of the U.S. Department of Energy by Lawrence Livermore National Laboratory under Contract DE-AC52-07NA27344.}
\end{abstract}

\section{Introduction}
Computed Tomography (CT) reconstruction is the process of recovering the structure and density of objects from a series of x-ray projections, called sinograms. While traditional full-view CT is relatively easier to solve, the problem becomes under-determined in two crucial scenarios often encountered in practice -- (a) few-view: when the number of available x-ray projections is very small, and (b) limited-angle: when the total angular range is less than 180 degrees, as a result of which most of the object of interest is invisible to the scanner. These scenarios arise in applications which require the control of x-ray dosage to human subjects, limiting the cost by using fewer sensors, or handling structural limitations that restrict how an object can be scanned. When such constraints are not extreme, suitable regularization schemes can help produce artifact-free reconstructions. While the design of such regularization schemes are typically driven by \emph{priors} from the application domain, they are found to be insufficient in practice under both few-view and limited-angle settings.


In the recent years, there is a surge in research interest to utilize deep learning approaches for challenging inverse problems, including CT reconstruction~\cite{anirudh2017lose,dong2019sinogram,CTShen2019}. These networks implicitly learn to model the manifold of CT images, hence resulting in higher fidelity reconstruction, when compared to traditional methods such as Filtered Backprojection (FBP), or Regularized Least Squares (RLS), for the same number of measurements. While these continue to open new opportunities in CT reconstruction, they rely of directly inferring mappings between sinograms and the corresponding CT images, in lieu of regularized optimization strategies. However, the statistics of sinogram data can vary significantly across different scanner types, thus rendering reconstruction networks trained on one scanner ineffective for others. Furthermore, in practice, the access to the sinogram data for a scanner could be restricted in the first place. This naturally calls for entirely image-domain methods that do not require access to the underlying measurements. In this work, we focus on the limited-angle scenario, which is known to be very challenging due to missing information. Instead of requiring sinograms or scanner-specific representations, we pursue an alternate solution that is able to directly work in the image domain, with no pairwise (sinogram--image) training necessary. 

To this end, we advocate the use of generative adversarial networks (GANs) \cite{GANGoodfellow} as image manifold priors. GANs have emerged as a powerful, unsupervised technique to parameterize high dimensional image distributions, allowing us to sample from these spaces to produce very realistic looking images. We train the GAN to capture the space of all possible reconstructions using a training set of clean CT images. Next, we use an initial seed reconstruction using an existing technique such as Filtered Back Projection (FBP) or Regularized Least Squares (RLS) and `clean' it by projecting it onto the image manifold, which we refer to as the GAN prior following \cite{shahICASSP2018}. Since the final reconstruction is always forced to be from the manifold, it is expected to be artifact-free. More specifically, this process involves sampling from the latent space of the GAN, in order to find an image that resembles the seed image. Though this has been conventionally carried out using projected gradient descent (PGD)~\cite{yeh2016semantic, shahICASSP2018}, as we demonstrate in our results, this approach performs poorly when the initial estimate is too noisy or has too many artifacts, which is common under extremely limited angle scenarios. 

Instead, our approach utilizes a recently proposed technique referred to as corruption mimicking, used in the design of MimicGAN \cite{anirudh2018mimicgan}, that achieves robustness to the noisy seed reconstruction through the use of a randomly initialized shallow convolutional neural network (CNN), in addition to PGD. By modeling the initial guess of this network as a random corruption for the unknown clean image, the process of corruption mimicking alternates between estimating the unknown corruption and finding the clean solution, and this alternating optimization is repeated until convergence, in terms of effectively matching the observed noisy data. The resulting algorithm is test time only, and can operate in an artifact-agnostic manner, i.e. it can clean images that arise from a large class of distortions like those obtained from various limited-angle reconstructions. Furthermore, it reduces to the well-known PGD style of projection, when the CNN is replaced by an identity function.

\section{Proposed Approach}
\label{sec:methods}
We restrict our study to parallel beam and fan beam types of scanners, that produces a CT reconstruction in a 2D slice-by-slice manner. The CT reconstruction problem, like most other inverse problems, can be written as: $X^* = \argmin_X \|\mathbf{A}(X)  - y\| + R(X)$, where $X \in \mathrm{R}^{d\times d}$ is the image to be reconstructed, $y \in \mathrm{R}^{v \times d}$ is the projection, referred to as a ``sinogram'', and $\mathbf{A}$ is the x-ray projection operator of the particular CT scanner. Here, the number of available x-ray projections is given by $v$, and the number of detector columns is given by $d$. Note, $\mathbf{A}(X)$ can be written as a matrix multiplication, but the matrix tends to be a sparse, very large matrix. Here, for simplicity we denote it as an operator acting on $X$.  Typically, a regularization function in the form of $R(X)$ is used to further reduce the space of possible solutions. In order to get a complete faithful reconstruction of $X$, the object must be scanned from a full $180^{\circ}$.  When the viewing angle is much lesser than $<<180^{\circ}$, most existing methods return an $X^*$ that is extremely corrupted by noise and missing edges, with little or no information of the original structure present. 

\noindent While several kinds of regularization functions have been used (for e.g. total variation and its variants), in this paper we advocate the use of $R(X)$ such that it forces $X$ to be from a known image manifold. We achieve this by using generative adversarial networks (GANs) \cite{GANGoodfellow}, which have emerged as a powerful way to represent image manifolds. In particular at test time, given a sinogram $y$, the problem can be formulated as 
\begin{equation}
\label{eq:manifold}
z^* = \argmin_{z\in \mathcal{U}(-1,1)} \|\mathbf{A}(\mathcal{G}(z))  - y\|,\mbox{~where~} \mathcal{G} \mbox{~is a pre-trained generator,}
\end{equation}
and finally, $X^* = \mathcal{G}(z^*)$ and can be solved using stochastic gradient descent. This has been referred to as a GAN prior \cite{shahICASSP2018} or a manifold prior \cite{CTShen2019} for inverse imaging. However, solving the equation of the form in \eqref{eq:manifold} is not always possible since one may not have access to the measurement model $\mathbf{A}$. A more accessible (yet different) form that does not require \textbf{A} to be known is given by: 
\begin{equation}
\label{eq:image_only}
z^* = \argmin_{z\in \mathcal{U}(-1,1)} \|\mathcal{G}(z)  - X_{RLS}\|,\mbox{~where~} X_{RLS} \mbox{~is an initial estimate.}
\end{equation}

In this work we obtain the initial estimate $X_{RLS}$ using regularized least squares (RLS) approach. As a result of \eqref{eq:image_only}, the quality of the final estimate largely depends on the quality of the initial reconstruction. Particularly, if the estimate is very noisy or poor, as is the case for limited angle CT, the optimization in \eqref{eq:image_only} can easily fail, especially when the loss is not robust to the type of corruption noise or distortion. In scenarios of interest in this paper, even a powerful regularizer such as the GAN prior can fail due to a poor initial estimate. In order to avoid this, we propose to use a recently proposed modification of the GAN prior, that performs better even with heavily distorted images. The process called \emph{corruption mimicking} was proposed in \cite{anirudh2018mimicgan}, was designed to improve the quality of projection onto the manifold under a variety of corruptions. 

\noindent \textbf{Corruption Mimicking and the Robust GAN Prior:} Let us suppose $X_{RLS} = f(X^*)$, where $f$ is an unknown distortion or corruption function, and $X^*$ is the unknown global optima to \eqref{eq:image_only}. Corruption Mimicking is the process of estimating both $X^*$ and $f$ simultaneously, using a shallow neural network to approximate $f$ with a few examples. As a result, we now modify \eqref{eq:image_only} as follows:
\begin{gather}
\label{eq:robust}
\hat{f}^*, z^* = \argmin_{z\in \mathcal{U}(-1,1); \hat{f}} \|\hat{f}(\mathcal{G}(z))  - X_{RLS}\|, \mbox{ where~} \hat{f} \mbox{~ is a shallow CNN}\\
\mbox{such that~} \hat{f}^* \approx f\mbox{~ and ~} \mathcal{G}(z^*) \approx X^*. 
\end{gather}

Equation \eqref{eq:robust} is solved using alternating optimization, where we first solve for the optimal $\hat{f}^*$ conditioned on the current estimate $X^*$, and repeat the process until convergence. Since we constrain $\hat{f}$ to be shallow, even as few as 100 samples are sufficient. In our setting, $\hat{f}$ contains 2 convolutional layers with ReLU activations, followed by a masking layer (pixel-wise multiplication). Finally, we also include a shortcut connection at the end to encourage it to learn identity \cite{anirudh2018mimicgan}. The GAN prior now becomes a special case of the Robust GAN prior, when $\hat{f} = \mathcal{I}$, the identity function. An appealing property of this technique is that it is \emph{corruption-agnostic} i.e., the same system can be reused to obtain accurate CT reconstructions across a wide variety of limited-angle settings.
\section{Experiments}
We test the effectiveness of the robust GAN prior by performing CT reconstruction of the MNIST \cite{lecun1998mnist} and Fashion-MNIST \cite{fashion-mnist} datasets. We first project these datasets into their projection space (sinograms), using a forward projection operation, to simulate the CT-scan process. While we consider a parallel beam scanner in these experiments, the methods and reported observations are applicable to other scanner types, since the proposed method operates directly in the image space. Next, we recover the images using the regularized least squares algorithm (RLS), which is commonly adopted in CT reconstruction. We emulate the limited-angle scenario by providing only a partial sinogram to RLS. We provide the resulting reconstruction as the input to the proposed algorithm.  

\noindent \textbf{Experimental Settings:} On both datasets, we train a standard DCGAN \cite{radford2015unsupervised} to generate images using the 60K training $28\times 28$ images. We run all our reconstruction experiments on a subset of the $10K$ validation set. Corruption-mimicking requires choosing 4 main hyperparameters \cite{anirudh2018mimicgan}: $T_1 = 15, T_2=15, \gamma_s=1e-2, \gamma_g=8e-2$ that control the number of iterations in the alternating optimization and learning rates (see section \ref{sec:methods} for details), these are kept fixed on both datasets, across all viewing angle settings. We observed the performance to be robust across a wide range of settings for these hyper-parameters. Finally, we compare the performance of the robust GAN prior against the standard GAN prior, without corruption-mimicking. In both cases, we run the latent space optimization for a total of $\sim 2500$ iterations, which typically only takes about 10 seconds on a P100 NVIDIA GPU. 

\begin{figure*}[!thb]
	\centering
	\includegraphics[trim={0 0 0 0},clip,width=0.95\linewidth]{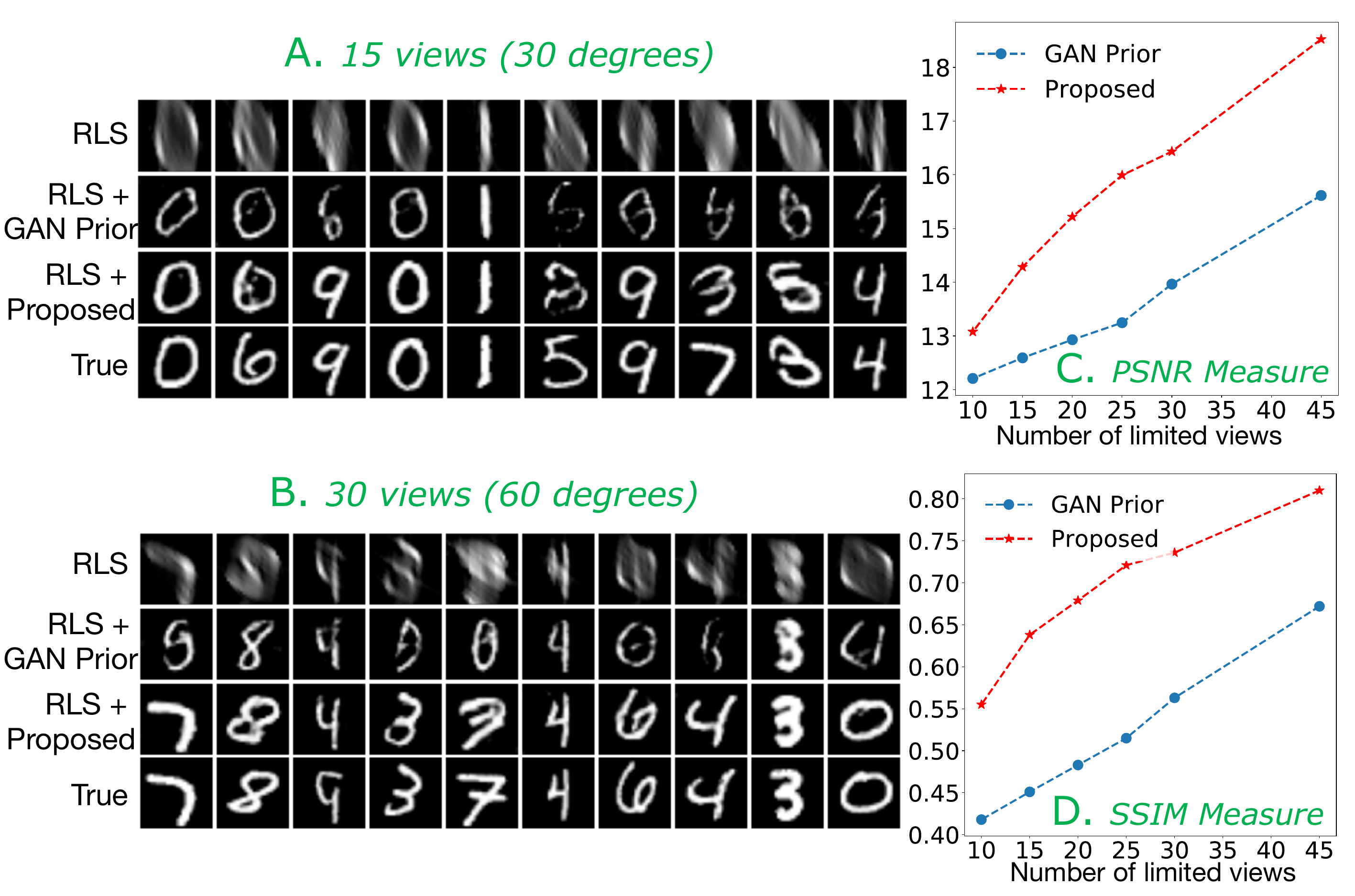}
	\caption{MNIST Dataset \cite{lecun1998mnist}: Given the RLS reconstruction, we improve them by projecting onto the image manifold using corruption mimicking\cite{anirudh2018mimicgan}. \textbf{(A)} Reconstructions from 30 degree limited angle. \textbf{(B)} Reconstructions from 60 degree limited angle. \textbf{(C)} PSNR metric for quality on 100 random test samples. \textbf{(D)} SSIM metric for quality. In all cases, we show the improvement obtained by using the robust GAN prior over a standard GAN projection. }
	\label{fig:mnist}
\end{figure*}

\begin{figure*}[!thb]
	\centering
	\includegraphics[trim={0 0 0 0},clip,width=0.95\linewidth]{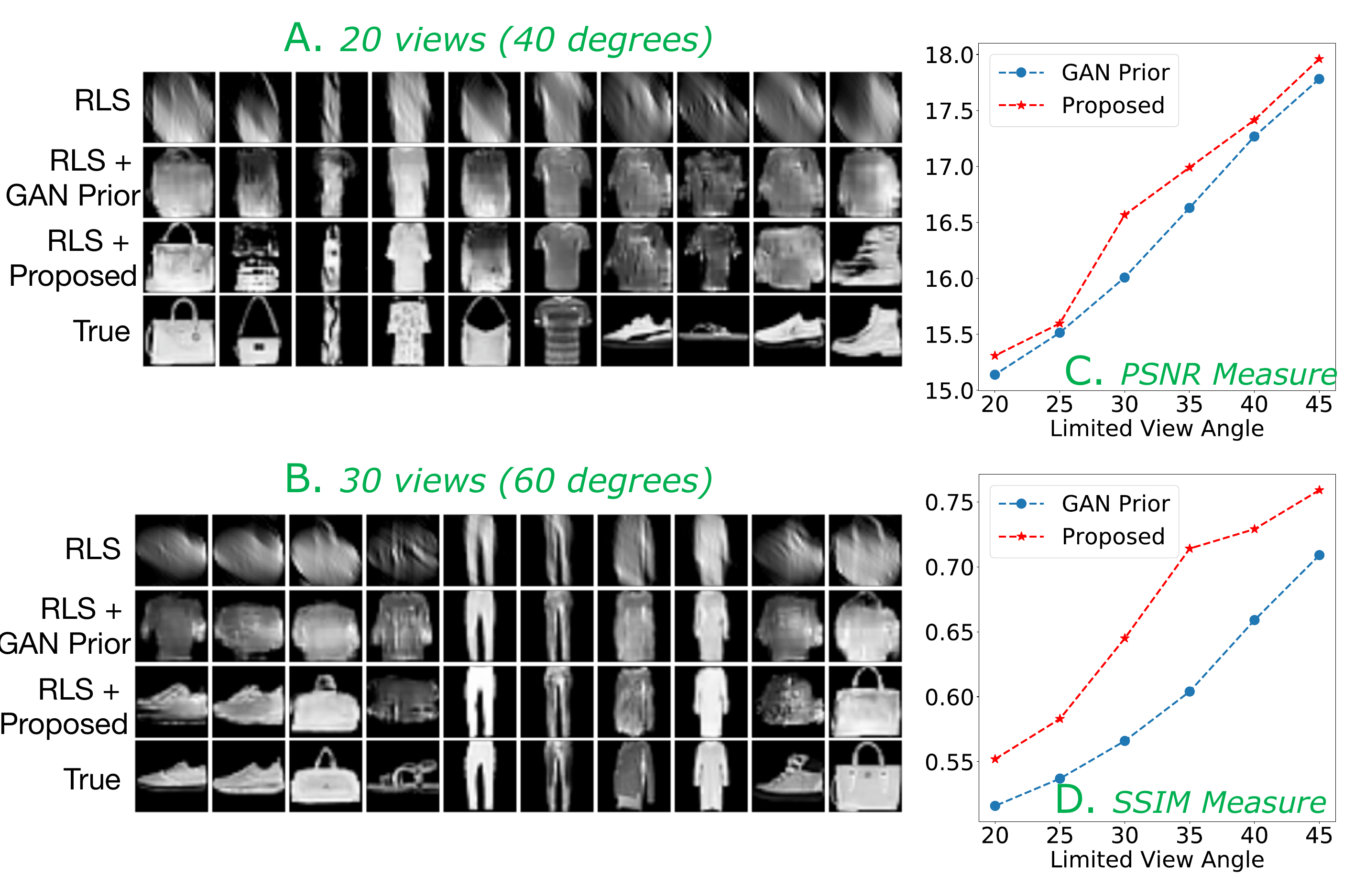}
	\caption{Fashion MNIST Dataset \cite{fashion-mnist} \textbf{(A):} Reconstructions from 40 degree limited angle. \textbf{(B):} Reconstructions from 60 degree limited angle. \textbf{(C):} PSNR metric for quality on 100 random test samples. \textbf{(D):} SSIM metric for quality. In all cases, we show the improvement obtained by using the robust GAN prior over a standard GAN projection. }
	\label{fig:fashion}
\end{figure*}

\noindent \textbf{Discussion and Results:} In figures \ref{fig:mnist}, \ref{fig:fashion}, we show qualitative and quantitative results obtained for both the MNIST and Fashion-MNIST datasets respectively. In both cases, we demonstrate significant improvements in recovering the true reconstruction compared to the vanilla GAN prior. It should be noted that a performance boost of nearly 4-5 dB on MNIST and 0.5-1dB on Fashion-MNIST are achieved with no additional information or data, but due to the inclusion of the robust GAN prior. Additionally, PSNR and SSIM tend to be uncorrelated with perceptual metrics in many cases, as perceptually poor reconstructions can be deceptively close in PSNR or SSIM. A potential fix in GAN-based reconstruction approaches is to compute error in the discriminator feature space as a proxy for perceptual quality.

\small{
\subsubsection*{Disclaimer}

 \noindent This document was prepared as an account of work sponsored by an agency of the United States government. Neither the United States government nor Lawrence Livermore National Security, LLC, nor any of their employees makes any warranty, expressed or implied, or assumes any legal liability or responsibility for the accuracy, completeness, or usefulness of any information, apparatus, product, or process disclosed, or represents that its use would not infringe privately owned rights. Reference herein to any specific commercial product, process, or service by trade name, trademark, manufacturer, or otherwise does not necessarily constitute or imply its endorsement, recommendation, or favoring by the United States government or Lawrence Livermore National Security, LLC. The views and opinions of authors expressed herein do not necessarily state or reflect those of the United States government or Lawrence Livermore National Security, LLC, and shall not be used for advertising or product endorsement purposes.}

\bibliographystyle{unsrt}  
\bibliography{refs}

\end{document}